\documentclass[conference, letterpaper]{IEEEtran}
\IEEEoverridecommandlockouts
\usepackage{cuted}
\usepackage{cite}
\usepackage{bm}
\usepackage{comment}
\usepackage{amsmath,amssymb,amsfonts}
\usepackage{algorithm}
\usepackage{lettrine}
\usepackage{graphicx}
\usepackage{subfigure}
\usepackage{multirow}
\usepackage{multicol}
\usepackage{algpseudocode}
\usepackage{diagbox}
\usepackage{graphicx}

\usepackage[letterpaper, left=0.625in, right=0.625in, bottom=1.02in, top=0.70in]{geometry}
\usepackage{textcomp}
\usepackage{xcolor}
\usepackage{booktabs}
\usepackage{makecell}
\usepackage{epstopdf}
\usepackage{array}

\newcommand{\be}{\begin{equation}}
\newcommand{\ee}{\end{equation}}
\newcounter{problem}
\newcommand{\problem}{\stepcounter{problem}(\text{P}_{\theproblem}) \quad}

\def\BibTeX{{\rm B\kern-.05em{\sc i\kern-.025em b}\kern-.08em
    T\kern-.1667em\lower.7ex\hbox{E}\kern-.125emX}}

\usepackage{hyperref}

\begin{document}

\title{Exploiting RIS Optimization Limits for Multi-User Beamforming and Signal Suppression}

\author{Subham Sabud, \emph{Student Member, IEEE}, 
Mengni Zhao, \emph{Student Member, IEEE}, \\ Chu Ma, \emph{Member, IEEE}, Suman Banerjee, \emph{Fellow, IEEE}, and Feng Ye, \emph{Senior Member, IEEE}
\thanks{This project is partially supported by the U.S. National Science Foundation under Grant ECCS-2336234.}
\thanks{
Subham Sabud, Chu Ma and Feng Ye (corresponding) are with the Department of Electrical and Computer Engineering, University of Wisconsin-Madison, Wisconsin, WI, USA. Subham and Mengni made equal contributions. Emails: \{sabud, mzhao262, chu.ma, feng.ye\}@wisc.edu. 
}
\thanks{
Mengni Zhao and Suman Banerjee are with the Department of Computer Science, University of Wisconsin-Madison, Wisconsin, WI, USA. Emails: \{mzhao262@wisc.edu, suman@cs.wisc.edu\}.
}
}

\maketitle

\begin{abstract}
This paper presents a unified framework for exploiting the boundaries of reconfigurable intelligent surfaces (RIS) joint optimization in multi-user wireless systems, where a single RIS accommodates diverse objectives.We first propose an adaptive gradient-scaling mechanism that accelerates the convergence of the underlying optimization algorithm while maintaining stable performance across varying channel and system parameters. The proposed mechanism enables the solver to reach a reasonably good solution rapidly without requiring manual tuning of step sizes or algorithmic hyperparameters when system inputs change. We then propose a low-complexity beamformer recovery method tailored for single-user scenarios, which circumvents the full matrix decomposition required by traditional approaches, thereby significantly reducing computational overhead. Building on these foundations, we develop an element allocation strategy that enables user-specific prioritization through assignment of RIS subsets. This is further extended by a modular add-drop mechanism that supports partial-panel optimization in general multi-user settings. The framework is evaluated across three representative scenarios: (i) signal amplification for all users, (ii) signal suppression for all users, and (iii) selective amplification and suppression. To characterize performance limits, we derive power trade-off boundaries using scalarized joint optimization, which closely align with Monte Carlo simulations. Our unified joint optimization method consistently yield solutions near these boundaries, confirming its near-optimality. Extensive simulations under realistic channel models demonstrate that the proposed approach outperforms conventional semidefinite relaxation techniques, offering a scalable and effective RIS control strategy for cooperative and competitive multi-user environments.
\end{abstract}

\begin{IEEEkeywords}
Reconfigurable Intelligent Surface (RIS), Joint Optimization, Signal Enhancement, Signal Jamming
\end{IEEEkeywords}

\section{Introduction}

Reconfigurable Intelligent Surfaces (RIS) have emerged as a transformative technology in wireless communications, offering the potential to reshape the propagation environment to meet the stringent demands of 6G and beyond networks \cite{pan2022reconfigurable, khan2025survey, umer2023ris}. By intelligently configuring their programmable elements, RIS can achieve a range of propagation control objectives, including beamforming, spatial multiplexing, and interference suppression, without relying on active RF chains, which are typically required in amplify-and-forward or decode-and-forward relays \cite{liu2020reconfigurable, shen2023multibeam, du2024interfsupp, YETENEK24RISVSAFDF, bjornson2019intelligent}. This passive nature enables RIS to significantly enhance both spectral and energy efficiency.
Recent research has focused on leveraging RIS for signal enhancement \cite{basar19singleueenhance,RISvsMIMO,cons&desForMultiUEs} and signal nulling \cite{LYU20jamsingleue,maliciousRIS}, primarily targeting a single user in MIMO systems. In these studies, other users are often treated as constraints or passive entities. However, a compelling and underexplored direction involves the joint use of RIS for both suppression and amplification across multiple users.

In this paper, we investigate a multi-user scenario where a single RIS panel is tasked with simultaneously enhancing the signal for selected users while suppressing it for others. This dual-objective configuration introduces a complex optimization landscape, particularly when user channels are heterogeneous and correlated. Our key contribution lies in explicitly incorporating multiple users into the objective function, enabling simultaneous optimization of distinct, user-specific requirements. This marks a shift from single-user-centric designs to a truly multi-objective control framework, significantly expanding the design space and introducing new theoretical and practical challenges. Several technical obstacles arise in addressing this joint suppression–amplification problem. First, as in single-user RIS scenarios, the optimization problem is inherently nonconvex, limiting the applicability of standard convex solvers. Second, the trade-off surface between competing objectives, such as maximizing one user’s signal power while minimizing another’s, exhibits sharp curvature and asymmetric sensitivity, complicating its characterization. Third, the Pareto frontier for joint RIS control remains poorly understood, especially in boundary regions where one objective dominates.

To tackle these challenges and explore the feasibility of managing multiple users with a single RIS, we propose a multi-user optimization framework for joint RIS reflection control. Central to our approach is a dynamic allocation scheme that adjusts the number of RIS elements assigned to each user, enabling a quantifiable and interpretable implementation of user-specific priorities. By mapping element allocation to weight coefficients in the joint objective function, we achieve a flexible realization of trade-offs between suppression and amplification. This structured decomposition allows us to approach the optimal configuration of an RIS across heterogeneous user demands.
Furthermore, we develop a boundary characterization strategy to delineate the achievable trade-off region. By formulating scalarized joint optimization problems, we derive boundary points that closely match those obtained via large-scale Monte Carlo simulations. Our proposed framework consistently yields solutions near these boundaries ($<0.5$~dB in 80\% cases), confirming its effectiveness in approaching optimal trade-offs. More extensive evaluations under realistic channel conditions further validate the performance of our approach, demonstrating consistent gains over our joint optimization methods. 
The main contributions of this work are summarized as follows:

\begin{itemize}
    \item This work proposes a unified RIS optimization framework for multi-user wireless systems, enabling flexible combinations of objectives such as simultaneous enhancement, suppression, or selective control of received power. User-specific weighting is achieved through dynamic allocation of RIS elements.
    
    \item An adaptive gradient-scaling mechanism is developed to enhance a model-based Deep Unfolding network, enabling faster convergence to a high-quality solution while maintaining stability across varying channel conditions and system parameters. By employing alternating optimization, the mechanism automatically adjusts update magnitudes at each layer, effectively resolving the scale mismatch issue inherent in passive RIS systems and eliminating the need for manual step-size tuning.
    
    \item A low-complexity beamformer recovery method is derived to bypass full matrix decomposition, significantly reducing computational overhead while maintaining performance. This method is incorporated into a modular element adjustment scheme that supports dynamic inclusion or exclusion of RIS elements. In multi-user scenarios, this scheme enables scalable joint optimization by controlling element-level contributions to each user's objective.

    \item A boundary characterization strategy is developed to construct the achievable trade-off region via scalarized joint optimization. The resulting boundaries closely align with Monte Carlo simulations, and the proposed solutions consistently approach these limits, demonstrating both theoretical rigor and practical effectiveness.

    \item Extensive evaluations under realistic channel models benchmark the proposed framework across cooperative, competitive, and adversarial multi-user scenarios. The results show consistent performance improvements over conventional semidefinite relaxation (SDR) methods, confirming the scalability and robustness of the approach.

\end{itemize}

\section{Related Work}\label{sec:related_work}

RIS has been widely studied for mitigating signal degradation in wireless environments with blockage or weak line-of-sight (LOS) links. By intelligently redirecting electromagnetic waves, RIS can create controllable reflected paths and improve coverage and signal strength. For example, Wu et al. demonstrated RIS-enabled virtual LOS links in non-line-of-sight (NLOS) millimeter-wave systems \cite{WU19enhance}, while Huang et al. studied RIS deployment for reducing severe urban path loss through optimized placement and phase configuration \cite{Huang19}.

Beyond coverage restoration, RIS has been used to improve SINR, sum rate, energy efficiency, and QoS in multi-user systems. Wu and Zhang proposed SDR-based joint active/passive beamforming for downlink multi-user MISO systems \cite{WU19enhance}; Guo et al. introduced fractional programming with alternating optimization for weighted sum-rate maximization under both perfect and imperfect CSI \cite{Guo20fracprog}; Ni et al. developed a multi-cell RIS-aided NOMA framework jointly optimizing transmit power, reflection coefficients, and decoding order \cite{NI20ITER}; and Mu et al. proposed STAR-RIS protocols supporting simultaneous transmission and reflection for improved energy efficiency and QoS \cite{mu22starris}. Since RIS phase design is non-convex due to unit-modulus constraints, common solution tools include SDR, alternating optimization, fractional programming, and manifold optimization. For instance, ElMossallamy et al. developed a Riemannian gradient method for RIS configuration in OFDM networks \cite{ElMossallamy21Manifold}. Recent data-driven approaches further use DRL for scalable RIS control, including TD3-based active/passive beamforming for MISO URLLC systems \cite{Hashemi23TD3DDPG} and LRCPPO-based optimization of STAR-RIS-assisted UAV NOMA emergency communication \cite{LEI24LRCPPO}.

Although RIS is mainly designed for performance enhancement, recent works have shown that malicious or misconfigured RIS panels can passively degrade legitimate transmissions by inducing destructive interference without active RF emission \cite{Hu24malrissecretkey,lin24malrisiot,Wang22illegalRIS}. Lyu et al. first demonstrated stealthy single-user RIS jamming through destructive reflection \cite{LYU20jamsingleue}, and Huang et al. extended this idea to multi-user downlink and uplink MISO systems, showing that even random RIS phase patterns can significantly reduce QoS \cite{Huang23jamMUMISOdownlink,Huang24jamMUMISOuplink}. To further exploit RIS suppression capability, SDR-based received-power minimization has been studied for targeted UEs \cite{maliciousRIS}, and Sabud et al. later proposed a DRL-based method for adaptive multi-user RIS jamming in downlink scenarios \cite{Sabud25MuJamRIS}.

Multi-user RIS control is especially challenging because a configuration beneficial to one user may degrade another, particularly under heterogeneous channels, frequencies, or operators \cite{multiUEchallenges}. Prior works have explored both constructive and adversarial settings, including sensing-SNR maximization/minimization under SINR and power constraints \cite{cons&desForMultiUEs}, selective phase-shift design for targeted degradation \cite{maliciousRIS}, and QoS-degrading weighted power optimization in MIMO systems \cite{RISvsMIMO}.In \cite{Sun2022}, a robust SINR-based beamforming framework is developed for multiuser IRS-assisted systems under incomplete CSI to guarantee outage-constrained performance for legitimate users. While effective for CSI uncertainty, it does not address selective signal amplification versus suppression, RIS element allocation, or trade-off boundary characterization. More recently, \cite{Wang2025} introduced intelligent omni-surfaces for simultaneous beamforming and active jamming, optimizing SINR-oriented metrics using additional surface capabilities. However, its objective is fixed and does not explore priority-driven multi-user trade-offs or the achievable boundary between conflicting amplification and suppression objectives.

In contrast, this work considers a passive RIS architecture and develops a unified multi-objective framework for selective signal amplification and suppression. By dynamically allocating RIS elements, employing modular element adjustment, and using scalarized optimization, the proposed method explicitly characterizes achievable trade-off boundaries and provides flexible priority control without active jamming or changes to the physical-layer architecture.


\section{System Model and Problem Formulation} \label{sec:pre methods}

\subsection{System Model}\label{subsec:system}

\begin{figure}[b]
\centering
\includegraphics[width=0.85\linewidth]{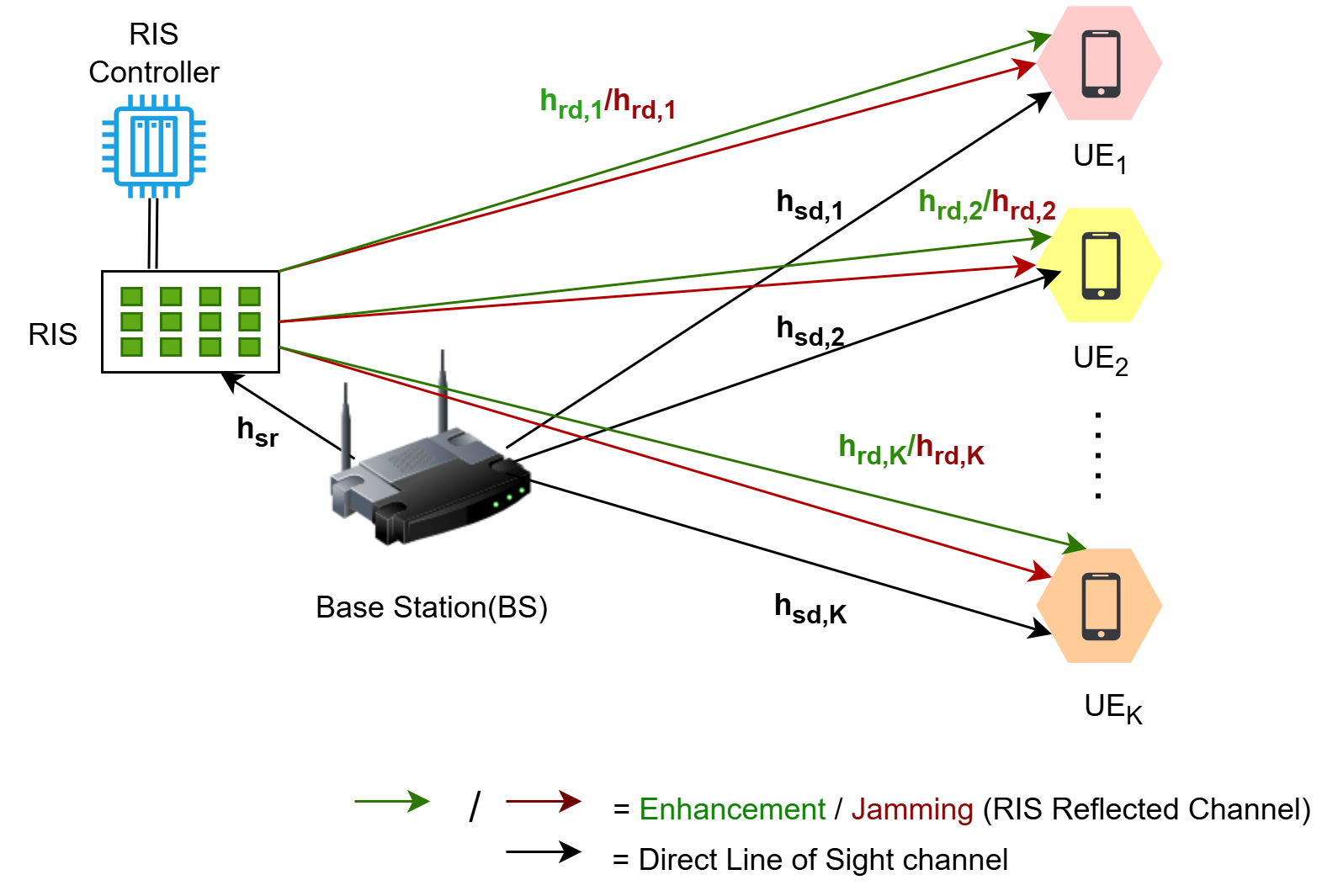}
\caption{Studied multi-user system model.}
\label{fig:System model}
\centering
\end{figure}

As demonstrated in Fig.~\ref{fig:System model}, the studied wireless communication system is composed of a single base station (BS) with $M$ antennas, a reconfigurable intelligent surface (RIS), and a set of $K$ user equipments (UEs), labeled as UE$_1$, UE$_2$,$\cdots$, UE$_K$. The RIS is deployed as a passive two-dimensional array composed of $ N$ reflecting elements, each capable of introducing a controllable phase shift to the incident electromagnetic wave. The collective phase response of the RIS is denoted by the vector $\boldsymbol{\Delta} = [\Delta_1, \Delta_2, \dots, \Delta_N]^\top \in \mathbb{C}^{N \times 1}$, where each element $\Delta_n = e^{j\theta_n}$ and $\theta_n \in [-\pi, \pi]$ represents the phase shift introduced by the $n$-th RIS element.
The channel model consists of three components for each UE: a direct line-of-sight (LOS) channel from the BS to the UE, and two cascaded links formed via the RIS.Specifically,  the BS-to-RIS channel is represented as $\mathbf{h}_{\mathrm{sr}} \in \mathbb{C}^{N \times M}$, while the RIS-to-UE$_k$ channel is given by $\mathbf{h}_{\mathrm{rd},k} \in \mathbb{C}^{N \times 1}$. The direct BS-to-UE$_k$ channel denoted as $\mathbf{h}_{\mathrm{sd},k} \in \mathbb{C}^{M \times 1}$. Taking into account both the direct and reflected paths, the composite effective channel between the BS and UE$_k$ becomes:
\begin{equation}
    \mathbf{h}_{\text{eff},k} = \mathbf{h}_{\mathrm{sd},k} + \hat{\mathbf{h}}_k \boldsymbol{\Delta},
\end{equation}
In this expression, cascaded channel can be expressed as $\hat{\mathbf{h}}_k = \mathbf{h}_{\mathrm{sr}} ^\top \mathrm{diag}(\mathbf{h}_{\mathrm{rd},k})$, where $\mathrm{diag}(\mathbf{h}_{\mathrm{rd},k})\in\mathbb{C}^{N\times N}$ represents the diagonal matrix associated with the RIS-to-UE$_k$ link. The base station transmits the symbol intended for UE~$k$ using a predetermined precoding vector $\mathbf{p}_k\in\mathbb{C}^{M}$. Under this configuration, the composite end-to-end channel between the BS and UE~$k$ can be modeled as an equivalent single-input single-output (SISO) channel, given as:
\begin{equation}
    h_{\text{eff},k} = \mathbf{p}_k^{\mathrm{T}}\mathbf{h}_{\text{eff},k}= \mathbf{p}_k^{\mathrm{T}}\mathbf{h}_{\mathrm{sd},k}
+ \mathbf{p}_k^{\mathrm{T}}\hat{\mathbf{h}}_k  \boldsymbol{\Delta}=h_{\mathrm{sd},k} + \mathbf{p}_k^{\mathrm{T}} \hat{\mathbf{h}}_k  \boldsymbol{\Delta},
\end{equation}
Based on the above signal model, the RIS-assisted received signal power at UE~$k$ can be expressed as
\begin{equation}
S_k \triangleq |h_{\mathrm{eff},k}|^2.
\end{equation}
To isolate the contribution of the RIS from the direct link, we further define the RIS-induced power variation as
\begin{equation}
P_k \triangleq S_k - |h_{\mathrm{sd},k}|^2. \label{eq:Pk_def_joint}
\end{equation}
The goal of the system is to configure the RIS phases such that the objectives of different UEs, which may include maximizing desired signal power or minimizing it, are jointly satisfied. The metrics $S_k$ and $P_k$ will be used throughout the paper as received-power-based performance measures to characterize signal amplification, suppression, and multi-user trade-offs.Since the precoding vectors $\{\mathbf{p}_k\}$
are fixed and the noise power is normalized, $S_k$ is equivalent
to the SNR at UE~$k$ up to a fixed positive scaling factor, and
no inter-user interference exists in the received signal model;
thus, optimizing $S_k$ is equivalent to optimizing
$\mathrm{SINR}_k$ under the considered system setup. CSI for all UEs is assumed perfectly known at the RIS controller, 
which is well-justified under the adopted Sionna RT ray-tracing methodology~\cite{10465179} 
that passively infers accurate spatial-CSI for all terminals without requiring their 
cooperation.

\subsection{Problem Formulation} \label{subsec: prob formulation}
\subsubsection{Vector-form Formulation}
For better illustration, the problem formulation begins with a single-user scenario. Assume that the RIS is configured to either minimize or maximize the received signal power at an arbitrary user $k$, the overall objective can be formulated as:

\begin{equation} \label{eq: vector form formu}
    \underset{{\boldsymbol{\Delta}}}{\text{min/max}} \quad  f_k(\boldsymbol{\Delta}) = |h_{\mathrm{eff},k}|^2 =  \left| h_{\mathrm{sd},k} + \mathbf{p}_k^{\mathrm{T}}\hat{\mathbf{h}}_k  \boldsymbol{\Delta} \right|^2.
\end{equation}
It is worth noting that $S_k$ is a physical received power metric under a fixed RIS configuration, while $f_k(\boldsymbol{\Delta})$ provides its functional representation with respect to the RIS variable $\boldsymbol{\Delta}$. For any given $\boldsymbol{\Delta}$, $f_k(\boldsymbol{\Delta})$ coincides with $S_k$, and the functional form enables optimization and boundary characterization.

The general multi-user optimization problem can be formulated in vector form as:
\begin{align}
\problem \quad \max_{\boldsymbol{\Delta}} \quad & \sum_{k \in \mathcal{K}_+} \alpha_k f_k(\boldsymbol{\Delta}) - \sum_{k \in \mathcal{K}_-} \alpha_k f_k(\boldsymbol{\Delta}) \label{eq: P1 prob formu: objective} \\
\text{s.t.} \quad 
& f_k(\boldsymbol{\Delta}) - \left| h_{\mathrm{sd},k}  \right|^2 \geq 0, \forall k \in \mathcal{K}_+, \label{eq:amp_constraint}\\
& f_k(\boldsymbol{\Delta}) - \left| h_{\mathrm{sd},k}  \right|^2 \leq 0, \forall k \in \mathcal{K}_-, \label{eq:supp_constraint}\\
& |\Delta_n| = 1, \quad \forall n = 1, \dots, N, \label{eq: vec_cons_1} \\
& \Delta_{N+1} = 1. \label{eq: P1 prob formu: last constraint}
\end{align}
where $\mathcal{K}_+$ and $\mathcal{K}_-$ denote the sets of users targeted for amplification and suppression, respectively; and $\alpha_k \ge 0$ are weight coefficients representing user priorities. The constraints in Eqs.~\eqref{eq:amp_constraint} and \eqref{eq:supp_constraint} impose minimal per-user guarantees with clear physical interpretations: users in $\mathcal{K}_+$ maintain at least the direct-path power, while users in $\mathcal{K}_-$ avoid RIS-induced enhancement.

\subsubsection{Equivalent Matrix-form Formulation}
To express this problem in a quadratic form, a Hermitian matrix $\mathbf{R}_k \in \mathbb{C}^{(N+1) \times (N+1)}$ is introduced to encapsulate the channel state information (CSI) for user $k$:
\begin{equation} \label{eq: R}
    \mathbf{R}_k = \begin{bmatrix} \hat{\mathbf{h}}_k^{\mathrm{H}} \hat{\mathbf{h}}_k & \hat{\mathbf{h}}_k^{\mathrm{H}} h_{\mathrm{sd},k} \\ h_{\mathrm{sd},k}^{\mathrm{H}} \hat{\mathbf{h}}_k & |h_{\mathrm{sd},k}|^2 \end{bmatrix}. 
\end{equation}
Eq.~\eqref{eq: vector form formu} is equivalent to
\begin{equation}
    \underset{\hat{\boldsymbol{\Delta}}}{\text{min/max}}  \quad\hat{\boldsymbol{\Delta}}^{\mathrm{H}} \mathbf{R}_k \hat{\boldsymbol{\Delta}}, \label{eq: single p}
\end{equation}
where 
\begin{equation} 
    \hat{\boldsymbol{\Delta}} = 
    \begin{bmatrix}
    \boldsymbol{\Delta} \\
    1
    \end{bmatrix}
    \in \mathbb{C}^{(N+1) \times 1},
\end{equation}
Since minimizing $\hat{\boldsymbol{\Delta}}^{\mathrm{H}} \mathbf{R}_k \hat{\boldsymbol{\Delta}}$ is equivalent to maximizing its negative, the multi-user problem integrating amplification and suppression objectives can be formulated as:
\begin{align}
    \problem \max_{\boldsymbol{\hat\Delta}} \quad & \boldsymbol{\hat\Delta}^\mathrm{H} \left( \sum_{k \in \mathcal{K}_+} \alpha_k \mathbf{R}_k - \sum_{k \in \mathcal{K}_-} \alpha_k \mathbf{R}_k \right) \boldsymbol{\hat\Delta} \label{eq: P2 prob formu: objective}\\
    \text{s.t.} \quad 
    & \boldsymbol{\hat\Delta}^\mathrm{H}  \mathbf{R}_k  \boldsymbol{\hat\Delta} - \left| h_{\mathrm{sd},k}  \right|^2 \geq 0, \forall k \in \mathcal{K}_+, \\
    &  \boldsymbol{\hat\Delta}^\mathrm{H}  \mathbf{R}_k  \boldsymbol{\hat\Delta} - \left| h_{\mathrm{sd},k}  \right|^2 \leq 0, \forall k \in \mathcal{K}_-,  \\
    & |\hat\Delta_n| = 1, \ \forall n \in \{1,\dots,N\}, \label{eq: norm = 1}\\
    & \hat{{\Delta}}_{N+1} = 1 \label{eq: P2 prob formu: last constraint}.
\end{align}
The direct resolution of this optimization problem is impeded by two fundamental obstacles. The first is the ambiguity of the weighting parameter $\alpha_k$, as it merely enforces a scalarized trade-off without explicitly reflecting the underlying performance disparities or fairness requirements across users. In practice, different values of $\alpha_k$ may lead to similar operating points, while small perturbations can result in disproportionate changes in user performance, making the priority interpretation ambiguous. This limitation motivates the adoption of more precise performance indicators for characterizing multi-user trade-offs.

Furthermore, from an optimization perspective, the multi-user objective function in Eq.~\eqref{eq: single p} becomes non-convex, in addition to the inherently non-convex unit-modulus constraint in Eq.~\eqref{eq: norm = 1}. As a result, relaxation techniques developed for the single-user case may not be directly applicable, which motivates the development of dedicated solution methods for both single-user and multi-user scenarios in the following sections.

\section{Approaches to Single UE Cases}\label{sec:single scheme}
The investigation begins with the most fundamental scenario, amplifying or suppressing the signal strength of a single target user. This simplified setting serves not only as a conceptual foundation for more complex multi-user signal control but also illustrates the core mechanism through which RIS manipulates the wireless propagation environment.

\subsection{Standard and Modified SDR Methods}\label{sec:baseline}


The single-user amplification problem follows directly from Eq.~\eqref{eq: vector form formu}. While gradient-based maximization methods can yield effective solutions, semidefinite relaxation (SDR) provides a stronger formulation and admits systematic extensions to multi-user scenarios. To be specific, we adopt an SDR-based linearization strategy and rewrite the objective as
\begin{align}   
    \hat{\boldsymbol{\Delta}}^{\mathrm{H}} \mathbf{R}_k \hat{\boldsymbol{\Delta}} 
    &= \mathrm{Tr} \left( \mathbf{R}_k \hat{\boldsymbol{\Delta}} \hat{\boldsymbol{\Delta}}^{\mathrm{H}} \right) \\
    &\triangleq \mathrm{Tr} \left( \mathbf{R}_k \boldsymbol{\Psi} \right),
    \quad \text{where } \boldsymbol{\Psi} = \hat{\boldsymbol{\Delta}} \hat{\boldsymbol{\Delta}}^{\mathrm{H}} .
\end{align}
The original optimization problem is then expressed as:
\begin{align}
    \problem \underset{{\boldsymbol{\Psi}}}{\text{maximize}}& \quad \mathrm{Tr}(\mathbf{R}_k \boldsymbol{\Psi}) \\
    \text{ s.t.}& \quad \mathrm{diag}(\boldsymbol{\Psi}) = 1, \label{eq: P3 first constraint}\\ 
    & \quad \boldsymbol{\Psi} \succeq 0, \\
    & \quad \mathrm{rank}(\boldsymbol{\Psi}) = 1. \label{eq: P3 last constraint, rank1constrain}
\end{align}
The rank-one constraint in Eq.~\eqref{eq: vec_cons_1} introduces non-convexity.
Conventional approaches address this via semidefinite relaxation (SDR) framework\cite{SDR}, which involves two steps: (i) relaxing the rank constraint to solve a convex problem, and (ii) recovering a rank-one solution by extracting the dominant eigenvector $\mathbf{v}{\mathrm{opt}}$ from the solution matrix $\boldsymbol{\Psi}{\mathrm{opt}}$, yielding:
\begin{equation}
    \boldsymbol{\Delta}_{\mathrm{opt}} = e^{j \angle \mathbf{v}_{\mathrm{opt}}(1:N)},
\end{equation}
and the final beamformer is $\hat{\bm{\Delta}}_{\mathrm{opt}}=[{\bm{\Delta}}_{\mathrm{opt}};1]$.



However, this recovery step requires full eigen-decomposition, incurring a computational cost of $\mathcal{O}(N^3)$, which is prohibitive for large-scale RIS systems. To overcome this, a modified SDR approach is proposed that avoids eigen-decomposition while maintaining comparable accuracy. By exploiting the structure of $\bm{\Psi}_{\mathrm{opt}}$, a high-quality rank-one approximation is reconstructed with significantly reduced complexity. Recall that $\bm{\Psi}$ is a Hermitian rank-one matrix defined as:
{\color{black}
 \begin{equation}
    \begin{split}
      \boldsymbol{\Psi} &= \hat{\boldsymbol{\Delta}} \hat{\boldsymbol{\Delta}}^{\mathrm{H}} \\
      &= \begin{bmatrix}
\Delta_1 \overline{\Delta}_1 & \Delta_1 \overline{\Delta}_2 & \Delta_1 \overline{\Delta}_3 & \cdots & \Delta_1 \overline{\Delta}_N & \Delta_1 \\
\Delta_2 \overline{\Delta}_1 & \Delta_2 \overline{\Delta}_2 & \Delta_2 \overline{\Delta}_3 & \cdots & \Delta_2 \overline{\Delta}_N & \Delta_2 \\
\vdots & \vdots & \vdots & \ddots & \vdots & \vdots \\
\Delta_N \overline{\Delta}_1 & \Delta_N \overline{\Delta}_2 & \Delta_N \overline{\Delta}_3 & \cdots & \Delta_N \overline{\Delta}_N & \Delta_N \\
\overline{\Delta}_1 & \overline{\Delta}_2 & \overline{\Delta}_3 & \cdots & \overline{\Delta}_N & 1
\end{bmatrix},
\end{split}
\end{equation}}
where $\overline{\Delta}_n$ denotes the complex conjugate of $\Delta_n$,  for $n = 1,\ldots,N$.
Evidently, the $(N{+}1)$-st column of $ \boldsymbol{\Psi}$ equals $\hat{\bm{\Delta}}$.
Despite relaxing the rank-one constraint,  $\bm{\Psi}_{\mathrm{opt}}$ often exhibits a dominant eigenvalue,
indicating near rank-one structure, suggesting a near-rank-one solution, as observed in \cite{MultiFunctionalRIS_ISAC}. 
Thus, instead of performing SDR recovery, the proposed method simply reads the last column of $\bm{\Psi}{\mathrm{opt}}$, extracts its phase, and normalizes the amplitudes:
\begin{equation}
\hat{\boldsymbol{\Delta}}_{\mathrm{opt}} = e^{j \cdot \angle(\Psi_{:, N+1})}, \quad n = 1, \ldots, N+1.
\end{equation}
This procedure has a computational complexity of only $\mathcal{O}(N)$, making it highly scalable and suitable for large-dimensional RIS applications and real-time deployment.



To approach the single-user suppression problem, we adopt a gradient projection method similar to the one used in~\cite{RISvsMIMO} to approximately convexify the optimization process. This approach is well-suited for optimizing quasi-convex objectives under unit-modulus constraints, which are inherent to RIS phase design.
The gradient of the objective with respect to the first $N$ entries of $\hat{\boldsymbol{\Delta}}$ is computed as:
\begin{equation}
\nabla f_k\!\left( \boldsymbol{\Delta} \right)
= \hat{\mathbf{h}}^{\mathrm{H}}
\left( h_{\mathrm{sd}} + \hat{\mathbf{h}}\, \boldsymbol{\Delta} \right).
\end{equation}
The descent update step is given by:
\begin{equation}
\boldsymbol{\Delta}^{(i)}
\leftarrow
\exp\!\left(
j \cdot
\angle\!\left(
\boldsymbol{\Delta}^{(i-1)}
-
\eta \nabla f_k\!\left( \boldsymbol{\Delta}^{(i-1)} \right)
\right)
\right), \label{eq: PGD single update}
\end{equation}
where ${i}$ denotes the iteration index, and $\eta$ is a fixed step size. This update preserves the phase structure of the gradient while reducing the objective value.
To enforce the unit-modulus constraint and fix the final element, the full vector is reconstructed as:
\begin{equation} 
    \hat{\boldsymbol{\Delta}}_{\mathrm{opt}} = 
    \begin{bmatrix}
    \boldsymbol{\Delta_{\mathrm{opt}}} \\
    1
    \end{bmatrix}.
\end{equation}
This iterative procedure is repeated until convergence.
Compared to penalty-based methods~\cite{maliciousRIS}, the gradient projection approach avoids reliance on convex solvers such as CVX, which may suffer from numerical instability in bi-criterion formulations. Instead, the projection-based updates maintain feasibility by design and demonstrate robust performance in practice.

\subsection{An AI-based Approach} \label{subsec: modelAI}

Although projected gradient descent (PGD) is widely used as a baseline solver for constrained optimization, it fundamentally relies on a fixed step size~$\eta$ chosen through domain knowledge. Since the optimal choice of~$\eta$ varies across problem instances, parameter ranges, and operating regimes, PGD lacks a universal update rule that can generalize across heterogeneous settings. This limits its applicability and transferability. To address this limitation, we adopt an AI-assisted approach inspired by~\cite{modelAI}, which embeds an iterative optimization algorithm into a trainable neural network through deep unfolding. In this framework, the conventional PGD update is generalized by replacing the fixed step size~$\eta$ with a layer-wise learnable parameter~$\eta^{(i)}$, enabling data-driven adaptation across iterations while preserving the underlying model structure.
Specifically, the RIS phase update at the $i$-th layer of the unfolded network is given by
\begin{equation}
\boldsymbol{\Delta}^{(i)}
\leftarrow
\exp\!\left(
j \cdot
\angle\!\left(
\boldsymbol{\Delta}^{(i-1)}
-
\eta^{(i)} \nabla f_k\!\left( \boldsymbol{\Delta}^{(i-1)} \right)
\right)
\right),
\end{equation}
where $i$ denotes the iteration (layer) index. Within this architecture, $\eta^{(i)}$ is treated as a trainable parameter optimized during learning, allowing each layer to adapt its update magnitude to the local geometry of the optimization landscape. The projection operator $\exp(j\angle(\cdot))$ ensures that the unit-modulus constraint on the RIS phase vector is strictly enforced at every iteration, thereby maintaining the feasibility of all updates.

However, directly applying this deep-unfolding approach to our problem introduces a critical challenge. Under practical RIS-assisted channel conditions, the cascaded channel coefficients $\hat{\mathbf{h}}_k$ typically exhibit very small magnitudes due to severe path loss. Consequently, the gradient $\nabla f_k(\boldsymbol{\Delta}^{(i)})$ can be orders of magnitude smaller than the phase $\boldsymbol{\Delta}$, creating a severe scale mismatch. This mismatch produces vanishingly small update steps under standard gradient-based unfolding, which in turn leads to numerically ineffective updates and significantly slows if not prohibits convergence.
To mitigate this issue, we introduce a scaling factor $\lambda \in \mathbb{R}$ and propose an alternating optimization procedure that updates $\lambda$ and $\eta$. Under this formulation, the update at iteration~$i$ becomes: 
\begin{equation}
\label{eq:scaled_update}
\boldsymbol{\Delta}^{(i)} 
= \boldsymbol{\Delta}^{(i-1)} 
- \eta^{(i)} \lambda^{(i)} 
\nabla f_k\!\left( \boldsymbol{\Delta}^{(i-1)} \right),
\end{equation}
For fixed $\eta$ in each layer, the optimization problem for $\lambda$ is: 
\begin{equation}
\label{eq:lambda_argmin}
\lambda^{(i)} 
= \arg\min_{\lambda}
\left| S_k^{(i-1)} 
- \eta^{(i)} \lambda c^{(i-1)} 
\right|^2,
\end{equation}
where
\begin{equation}
\label{eq:intermediate_signal}
S_k^{(i-1)} 
= h_{\mathrm{sd,k}} 
+ \mathbf{p}_k^{\mathrm{T}}\hat{\mathbf{h}}_k \boldsymbol{\Delta}^{(i-1)} ,
\end{equation}
\begin{equation}
\label{eq:gradient_direction}
c^{(i-1)} 
= \mathbf{p}_k^{\mathrm{T}}\hat{\mathbf{h}_k} 
\nabla f_k\!\left( \boldsymbol{\Delta}^{(i-1)} \right),
\end{equation}
with $c^{(i-1)} \in \mathbb{C}$ defined as an auxiliary variable to simplify the subproblem formulation.
Since~\eqref{eq:lambda_argmin} is an unconstrained convex quadratic program, the closed-form solution obtained via the first-order optimality condition is:
\begin{equation}
\label{eq:lambda_closed_form}
\lambda^{(i)}_{\text{exact}}
= \frac{
\Re \left\{ S_k^{(i-1)} \left( c^{(i-1)} \right)^*\right\}
}{
 \eta ^{(i)}\left| c^{(i-1)} \right|^2
}.
\end{equation}
However, substituting $\lambda^{(i)}_{\mathrm{exact}}$ into the update rule reveals a fundamental issue: the factor $\eta^{(i)}$ cancels with the $1/\eta^{(i)}$ term in~\eqref{eq:lambda_closed_form}, effectively removing $\eta^{(i)}$ from the computational graph and preventing it from being trained via backpropagation. This collapses the layer into a fixed, non-adaptive iteration. To prevent this cancellation and retain trainability, we adopt a decoupling strategy by computing $\lambda^{(i)}$ assuming a nominal unit step size (i.e., setting $\eta^{(i)} = 1$ temporarily). In this way, $\lambda^{(i)}$ captures curvature information of the objective function, while $\eta^{(i)}$ remains a learnable relaxation parameter that fine-tunes the update magnitude.
Additionally, $\lambda^{(i)}$ must be restricted to non-negative values to avoid inadvertently reversing the update direction, since the descent direction is entirely determined by $\nabla f_k$. 
as seen in the update rule in Eq.~\eqref{eq:scaled_update}, 
Since the scalar $\lambda^{(i)}$ acts as a multiplicative coefficient to the gradient term in the update rule (Eq.~\eqref{eq:scaled_update}), directly imposing a hard threshold, such as $\max(0, \lambda^{(i)})$, would nullify the update whenever the estimate becomes negative, resulting in $\boldsymbol{\Delta}^{(i)} = \boldsymbol{\Delta}^{(i-1)}$ and halting the optimization process. To ensure both numerical stability and non-negativity, we instead apply an absolute-value operator, leading to the adopted scaling factor:
\begin{equation}\label{eq:scaling factor}
\lambda^{(i)} = \left| \frac{\mathfrak{R}\left\{S_k^{(i-1)} \left( c^{(i-1)} \right)^* \right\}}{|c^{(i-1)}|^2}\right|. 
\end{equation}
The set of step sizes $\boldsymbol{\eta} = {\eta^{(1)}, \dots, \eta^{(L)}}$ then constitutes the trainable parameters of the network. These parameters are optimized end-to-end via backpropagation to minimize the loss function, defined as $f_k(\boldsymbol{\Delta})$ for minimization tasks and $-f_k(\boldsymbol{\Delta})$ for maximization tasks.

Finally, the iterative procedure is unfolded into an $I$-layer deep network, where each layer corresponds to one iteration of the underlying algorithm and is equipped with its own learnable step size~$\eta^{(i)}$. The complete algorithmic steps are summarized in Alg.~\ref{alg:deep_unfolding}.

\begin{algorithm}[ht!]
\caption{Adaptive-scaling deep unfolding network}
\label{alg:deep_unfolding}
\begin{algorithmic}[1]

\Require Training dataset $\mathcal{D}$, batch size $B$, learning rate $\gamma$, number of epochs $E$
\Ensure Optimized trainable step-size parameters $\boldsymbol{\eta} = \{\eta^{(1)}, \dots, \eta^{(I)}\}$

\State Initialize $\eta^{(i)}$, e.g., to $1$ for all $i$.

\For{$\text{epoch} = 1$ to $E$}
    \For{each batch $\mathcal{B} \subset \mathcal{D}$}

        \State \textbf{Forward propagation:}
        \State Initialize RIS phase vector $\boldsymbol{\Delta}^{(0)}$ randomly

        \For{$i = 1$ to $I$}  \Comment{Unfolding layer $i$}

            \State Compute gradient $\mathbf{g}^{(i-1)} = \nabla f\!\left( \boldsymbol{\Delta}^{(i-1)} \right)$.

            \State \textit{(1) Model-based scaling step ($\lambda$-update):}
            \State Compute scaling factor $\lambda^{(i)}$ using Eq.~\eqref{eq:scaling factor}

            \State \textit{(2) Data-driven update step ($\eta$-update):}
            \State Update intermediate vector:
            \[
            \mathbf{Z}^{(i)} 
            = \boldsymbol{\Delta}^{(i-1)}
            - \eta^{(i)} \lambda^{(i)} \mathbf{g}^{(i-1)}.
            \]

            \State \textit{(3) Projection:} Enforce unit-modulus constraint
            \[
            \boldsymbol{\Delta}^{(i)} = \exp\!\left( j\, \angle\, \mathbf{Z}^{(i)} \right)
            \]

        \EndFor

        \State \textbf{Loss computation:}
        \State Compute cumulative loss 
        \[
        \mathcal{L} = \sum_{i=1}^{I} \log_2(i+1) \cdot \mathcal{L}^{(i)}
        \]

        \State \textbf{Backward propagation:}
        \State Compute gradients $\nabla_{\boldsymbol{\eta}} \mathcal{L}$ and update parameters.
        

    \EndFor
\EndFor

\State \Return optimized step-size parameters $\boldsymbol{\eta}$

\end{algorithmic}

\end{algorithm}

\section{Joint Optimization for Multi-User Cases}\label{sec:joint_opt}

While the multi-user optimization problem could theoretically be formulated as a weighted-sum maximization (as in Eq.~\eqref{eq: P2 prob formu: objective}--~\eqref{eq: P2 prob formu: last constraint}, such a ``one-time'' global optimization approach often acts as a black box. It produces a final configuration without revealing the underlying trade-offs or the dynamic competition for RIS resources among users. In practical scenarios involving conflicting objectives (e.g., jamming vs. amplification), it is crucial to make the resource allocation process transparent and explainable, rather than merely presenting a static result.
To operationalize this transparent resource contention, we formulate the interaction among users as a strategic competition for finite RIS resources.

The joint RIS phase design problem is modeled for $K$ users, each aiming to either maximize or minimize their RIS-induced power variation. The optimization proceeds iteratively, with the RIS elements  partitioned among users based on their optimization outcomes.By decomposing the global problem into sequential sub-problems, our approach explicitly models the resource contention process, allowing users to progressively "claim" RIS elements based on their immediate utility gains. This not only renders the trade-off process transparent but also provides fine-grained control over the priority-driven allocation.


\textbf{Fixing Stage:} At the beginning of each iteration, the set of unfixed elements $\mathcal{F}^c$ is identified. Each user $k$ optimizes over these elements to either maximize or minimize $S_k$ according to their objective. If no elements are fixed, a full-panel SDR formulation is used; otherwise, a partial-panel SDR formulation is applied with an updated effective direct term.

\textbf{Effective Channel Definitions:}
Let $\mathcal{F}$ denote the set of indices for the currently fixed elements, and $\mathcal{U}$ denote the set of remaining unfixed elements. The core idea is to treat the signal reflection from $\mathcal{F}$ as a static component of the propagation environment, effectively merging it into the direct path.

The effective direct channel for user $k$ is denoted as $h_{sd,k}^{\text{eff}}$, aggregating the physical direct channel with the interference (or assistance) introduced by the fixed RIS elements:
\begin{equation}
    h_{sd,k}^{\text{eff}} = h_{sd,k} + \mathbf{p}_{k, \mathcal{F}}^\mathrm{T} \hat{\mathbf{h}}_{k, \mathcal{F}} \boldsymbol{\Delta}_{\mathcal{F}}, \label{eq:heff_update_joint}
\end{equation}
where $\boldsymbol{\Delta}_{\mathcal{F}}$ is the vector of fixed phase shifts, and $\hat{\mathbf{h}}_{k, \mathcal{F}}$ denotes the sub-matrix of the cascaded channel $\hat{\mathbf{h}}_k$ corresponding to the columns indexed by $\mathcal{F}$. Simultaneously, for the optimization of the remaining elements, we define the active cascaded channel $\hat{\mathbf{h}}_{k, \mathcal{U}}$ as the sub-matrix of $\hat{\mathbf{h}}_k$ containing only the columns indexed by $\mathcal{U}$.

\textbf{Generalized Optimization Formulation:}
With these definitions, the optimization for the unfixed subset $\mathcal{U}$ can be formulated as a generalized problem that mirrors the structure of the original single-user problem in Eq.~\eqref{eq: vector form formu}. The objective is to find the phase vector $\boldsymbol{\Delta}_{\mathcal{U}}$ that minimizes or maximizes the received power based on the effective channel conditions:
\begin{equation}
    \min_{\boldsymbol{\Delta}_{\mathcal{U}}} / \max_{\boldsymbol{\Delta}_{\mathcal{U}}} \quad f_k^{\text{sub}}(\boldsymbol{\Delta}_{\mathcal{U}}) = \left| h_{sd,k}^{\text{eff}} + \mathbf{p}_k^\mathrm{T}  \hat{\mathbf{h}}_{k, \mathcal{U}} \boldsymbol{\Delta}_{\mathcal{U}} \right|^2,
\end{equation}
subject to the unit-modulus constraints on $\boldsymbol{\Delta}_{\mathcal{U}}$. This sub-problem maintains the exact same functional form as (5) but with reduced dimensionality ($|\mathcal{U}| \leq N$), allowing it to be efficiently solved using the adaptive deep unfolding network described in Section~\ref{subsec: modelAI}. It is worth noting that this formulation unifies both full-panel and partial-panel optimization. In the initial iteration where $\mathcal{F} = \emptyset$, the effective direct channel reduces to the physical direct channel ($h_{sd,k}^{\text{eff}} = h_{sd,k}$), and $\hat{\mathbf{h}}_{k, \mathcal{U}}$ restores to the full cascaded channel $\hat{\mathbf{h}}_k$, thereby recovering the global optimization problem.
    

From the adaptive deep unfolding output, the user selects a subset of $M_k$ elements from $\mathcal{F}^c$ whose phase settings yield the largest improvement in $P_k$ as defined in Eq.~\eqref{eq:Pk_def_joint}. The fixed set $\mathcal{F}$ is updated by fixing phases in $\hat{\boldsymbol{\Delta}}$.  The effective direct term for user $k$ is updated as Eq.~\eqref{eq:heff_update_joint}.

\paragraph{Changing Stage} 
Once all $N$ elements are fixed, users who fail to meet their performance thresholds may attempt to reassign elements from others. For an unsatisfied user $k$, the target phase is computed as:
\begin{equation}
\theta_k = \angle(h^{\text{eff}}_{sd,k}) + (1 - I_{\max})\pi,
\end{equation}
where $I_{\max} = 1$ for maximization and $I_{\max} = 0$ for minimization. For each element $i$ assigned to another user, the angular mismatch is evaluated:
\begin{equation}
\delta_i^{(k)} = \left| \angle \left( [\mathbf{p}_k]_i[\mathbf{h}_k]_i \Delta_i \right) - \theta_k \right| 
\end{equation}
If $\delta_i^{(k)}$ exceeds a threshold $\delta_{\mathrm{thresh}}$, the element is reassigned to user $k$ with the updated phase:
\begin{equation}
\Delta_i \leftarrow \exp \left( j \left[ \theta_k - \angle \left( [\mathbf{p}_k]_i[\mathbf{h}_k]_i \right) \right] \right) 
\label{eq:delta_reassign_joint}
\end{equation}
and the effective direct channels for all users are updated:
\begin{equation}
h_{sd,l}^{\text{eff}} \leftarrow h_{sd,l} + \mathbf{p}_k^\mathrm{T} \mathbf{h}_{l,\mathcal{F}} \boldsymbol{\Delta}_{\mathcal{F}}, \quad \forall l = 1, \dots, K 
\end{equation}



The overall muilti-user optimization process is summarized in Alg.~\ref{alg:game_opt}. Although the proposed formulation adopts a game-theoretic viewpoint to model the competing objectives among users, the resulting optimization is not performed in a distributed manner by the users themselves. Instead, the joint RIS optimization, including the iterative updates and element reallocation steps described in Alg.~\ref{alg:game_opt}, is executed centrally by the network controller (e.g., the base station or a centralized RIS controller). The term ``user-wise optimization'' is used to describe how individual users’ objectives are incorporated into the joint problem formulation, rather than to imply decentralized decision-making.

\begin{algorithm}[ht!]
\caption{Joint RIS Optimization for $K$ Users}
\label{alg:game_opt}
\begin{algorithmic}[1]

\State \textbf{Initialization:}
\State $\hat{\boldsymbol{\Delta}} \leftarrow \mathbf{1}_{N+1}$; $\mathcal{F} \leftarrow \emptyset$; \quad
$h^{\mathrm{eff}}_{sd,k} \leftarrow h_{sd,k}$ for all $k \in \{1,\dots,K\}$;

\While{not all users satisfy their objectives}

    \For{$k = 1$ to $K$}

        \State {\textbf{Step 1: Generalized Optimization}}
        \State {Solve the generalized sub-problem over unfixed elements $\mathcal{F}^c$ using  the adaptive deep-unfolding network (based on effective channel $h^{\mathrm{eff}}_{sd,k}$).}

        \State{\textbf{Step 2: Element Selection and Fixing}}
        \State Select $M_k$ elements from $\mathcal{F}^c$ that yield the largest improvement in $P_k$ (Eq.~\eqref{eq:Pk_def_joint});
        \State Fix selected phases in $\hat{\boldsymbol{\Delta}}$ and update the set $\mathcal{F}$;

        \State {\textbf{Step 3: Environment Update}}
        \State Update $h^{\mathrm{eff}}_{sd,k}$ using Eq.~\eqref{eq:heff_update_joint};
        \State Compute $P_k$ using Eq.~\eqref{eq:Pk_def_joint};

    \EndFor

    \If{$|\mathcal{F}| = N$ { and some users remain unsatisfied}}

        \For{each unsatisfied user $k$}

            \For{each fixed element $i \in \mathcal{F}_j$, $j \neq k$}

                \State Compute $\delta_i^{(k)}$;
                
                \If{$\delta_i^{(k)} > \delta_{\mathrm{thresh}}$};
                    \State Reassign $\Delta_i$ using Eq.~\eqref{eq:delta_reassign_joint};
                    \State Update $h^{\mathrm{eff}}_{sd,l}$ for all $l$ using Eq.~\eqref{eq:heff_update_joint}
                \EndIf

            \EndFor

        \EndFor

    \EndIf

\EndWhile

\State \textbf{Return:} $\hat{\boldsymbol{\Delta}}$

\end{algorithmic}
\end{algorithm}

\section{Boundary Approximation via Scalarized Optimization}\label{sec:Boundary determination}


This section characterizes the performance boundary obtained via scalarized optimization for the proposed joint RIS control framework. By assigning weights to different user objectives and sweeping the weight vector, the multi-objective problem defined in Section~\ref{subsec: prob formulation} is converted into a sequence of single-objective optimizations, allowing different operating points on the achievable power region to be obtained, thereby approximating the Pareto boundary of the multi-user power control trade-off. To efficiently solve the resulting problems, the proposed framework combines SDR-based formulations with gradient-based optimization, exploiting the complementary strengths of both approaches. Although a formal proof of global optimality is intractable, the resulting boundary is empirically tight and closely matches Monte Carlo sampling results, indicating near-optimal performance in practice.

\subsection{Signal Amplification of all UEs} \label{subsec: all amp}
The problem formulation corresponds to the general multi-user problem defined in Eq~\eqref{eq: P1 prob formu: objective}--\eqref{eq: P1 prob formu: last constraint}. By introducing a composite channel correlation matrix $\mathbf{R}$, defined as the convex combination of individual user channel matrices:
\begin{equation}
    \mathbf{R} = \sum_{k=1}^{K} \alpha_k \mathbf{R}_k, \quad \text{with } \sum \alpha_k = 1, \alpha_k \ge 0
\end{equation}
the original multi-user objective can be reformulated into a trace maximization problem equivalent to the single-user case derived in Eq.~\eqref{eq: P2 prob formu: objective}:
\begin{equation}    
    \underset{{\boldsymbol{\Psi}}}{\max} \quad \mathrm{Tr}(\mathbf{R} \boldsymbol{\Psi}),
\end{equation}
with all the constraints identical to those in Eqs.~\eqref{eq: norm = 1} and \eqref{eq: P2 prob formu: last constraint}. Consequently, the optimal solution can be efficiently obtained by directly applying the low-complexity beamformer recovery method (Section~\ref{sec:baseline}). By traversing $\boldsymbol{\alpha}$, we trace the Pareto frontier for joint signal amplification.

\subsection{Signal Suppression of all UEs} \label{subsec: all jamming}

To approximate the lower boundary of the achievable power region, we consider the specific case of Problem ($\mathrm{P_1}$) where all users are targeted for suppression (i.e., $k \in \mathcal{K}_- $ for all $k$). This yields the following scalarized vector-form minimization problem:
\begin{equation} \label{eq: sup_vector}
\min_{\boldsymbol{\Delta}} \sum_{k=1}^{K} \alpha_k f_k(\boldsymbol{\Delta}), \quad \text{s.t. Eqs. } \eqref{eq:supp_constraint}, \eqref{eq: vec_cons_1}, \eqref{eq: P1 prob formu: last constraint}.
\end{equation}
A key property here is that the weighted sum of positive semi-definite (PSD) matrices remains PSD. Consequently, the objective function in the equivalent matrix form preserves the convex quadratic structure.
Therefore, instead of resorting to high-complexity relaxation, we can directly solve the vector-form problem \eqref{eq: sup_vector} using our AI-based approached (Section~\ref{subsec: modelAI}). This allows us to efficiently trace the boundary by traversing $\alpha_k$.



\subsection{Joint Amplification and Suppression} \label{subsec: joint jam and amp}
We next consider the general case where amplification and suppression objectives coexist across different users. For this hybrid objective, we evaluate two solution strategies derived from the formulations in Section~\ref{subsec: prob formulation}. Specifically, the matrix-form representation (Eq.~\eqref{eq: P1 prob formu: objective}--\eqref{eq: P1 prob formu: last constraint}) leads to a scalarized SDR formulation, while the vector-form representation (Eq.~\eqref{eq: P2 prob formu: objective}-- \eqref{eq: P2 prob formu: last constraint}) results in a scalarized PGD optimization. Since the resulting problem is non-convex, it is generally difficult to determine a priori which formulation leads to better solutions. Although scalarization cannot guarantee recovery of the entire Pareto frontier for non-convex problems, it remains a widely adopted technique for exploring convex trade-off boundaries in multi-objective optimization~\cite{miettinen1999nonlinear}.Therefore, both scalarized SDR and scalarized PGD are evaluated to trace the boundary of the achievable region.

\begin{table*}[t]
\centering
\caption{Evaluation results of single-user cases. ($^\dagger$ Using manually tuned optimal step sizes for each specific location.$^\ddagger$ Using a fixed step size of $\eta = 1,500,000$ tuned for location (3,0.5).)}
\label{tab:result comparison all locations}
\begin{tabular}{c|c|c|c|c|c|c}
\toprule
Location & Metric & {SDR} & {PGD (Optimized $\eta$)$^\dagger$} & PGD (Fixed $\eta$)$^\ddagger$ & Model-based DU & Ours (Single) \\
\midrule
\multicolumn{7}{c}{RIS optimization performance comparison (in dB)}\\
\midrule
(3,0.5) & Min & 9.2151 & -155.0119 & -155.0119 & 0.1031 & -119.8522 \\
        & Max & 13.7845 & 13.7845  & 13.7845 & 0.0535 & 13.7796 \\
\hline
(4,1) & Min & 9.4761 & -170.9772 & -147.2793 & -0.2432 & -74.2307 \\
      & Max & 13.9397 & 13.9397 & 13.9397 & 0.5812 & 11.9931 \\
\hline
(7.5,5.5) & Min & 9.6394 & -166.9773 & -27.0476 & 0.1844 & -114.1373 \\
          & Max & 14.0377 & 14.0361 & 13.9314 & 0.1019 & 14.0357 \\
\hline
(4.8,1.3) & Min & 9.6026 & -162.7014 & -100.2794 & 0.3326 & -113.9242 \\
          & Max & 14.0156 & 14.0155 & 14.0155 & 0.2093 & 14.0141 \\
\bottomrule
\end{tabular}
\end{table*}

\section{Evaluation Results}\label{sec:evaluation}

\subsection{Evaluation Settings}\label{subsec: Eva Settings}

We evaluate the proposed RIS-aided joint optimization framework in a $5$~GHz indoor wireless environment consisting of one BS, one RIS, and multiple UEs. The BS is located at $(0,0,0)$~m and the RIS at $(-0.3,0,0)$~m, representing a practical wall-mounted deployment near the transmitter. Unless otherwise stated, the RIS contains $N=100$ reflecting elements, the BS transmit power is fixed at $20$~mW, and the noise power is normalized to isolate the effect of RIS phase optimization.

CSI for the BS--RIS--UE and BS--UE links is generated using the Sionna Ray Tracing framework~\cite{10465179}, which captures both LOS and NLOS multipath components in indoor environments. The generated CSIs are then fixed and used as inputs for all optimization experiments, separating physical channel modeling from algorithmic evaluation. Sionna RT simulations are performed on a workstation with an Intel Core i9-13900K CPU, 125~GB RAM, and an NVIDIA RTX 4090 GPU running Ubuntu 22.04.5 LTS, while all optimization, boundary characterization, and performance evaluations are conducted on a separate Intel Core i5-11300H CPU platform with 16~GB RAM.

The evaluation covers both single-user and multi-user scenarios, including signal enhancement, signal suppression, selective enhancement/suppression, and achievable boundary characterization. Unless otherwise specified, all reported power values follow Eq.~\eqref{eq:Pk_def_joint}, i.e., the RIS-induced power variation relative to the direct-link-only case. The proposed methods are benchmarked against SDR-based techniques and brute-force Monte Carlo sampling whenever applicable.





\subsection{Evaluations on Single-User Optimization Limits}
\begin{figure}[htbp]
    \centering
    \subfigure[User located at (4.8, 1.3), Min.\label{fig:iter_4813_min}]{%
        \includegraphics[width=.48\linewidth]{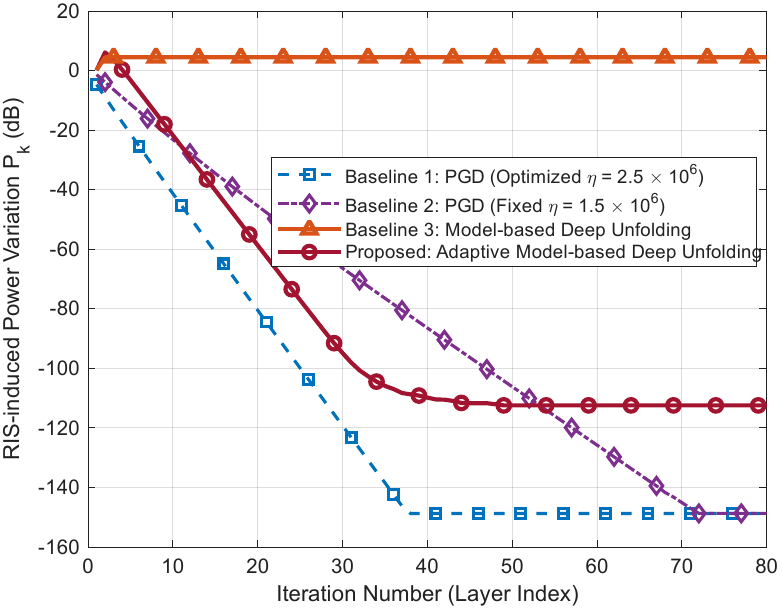}%
    }%
    \hfill
    \subfigure[User located at (7.5, 5.5), Min.\label{fig:iter_7555_min}]{%
        \includegraphics[width=.48\linewidth]{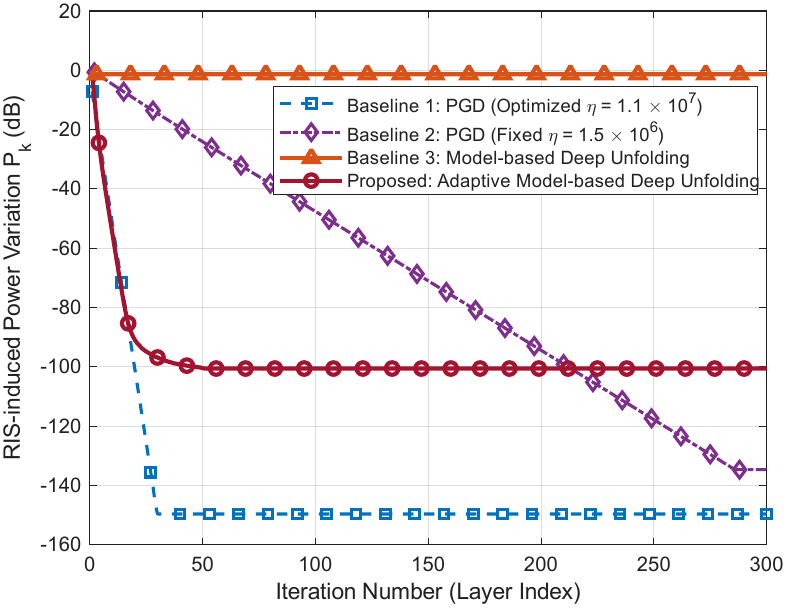}%
    }%

    \vspace{0.5em}

    \subfigure[User located at (4.8, 1.3), Max.\label{fig:iter_4813_max}]{%
        \includegraphics[width=.48\linewidth]{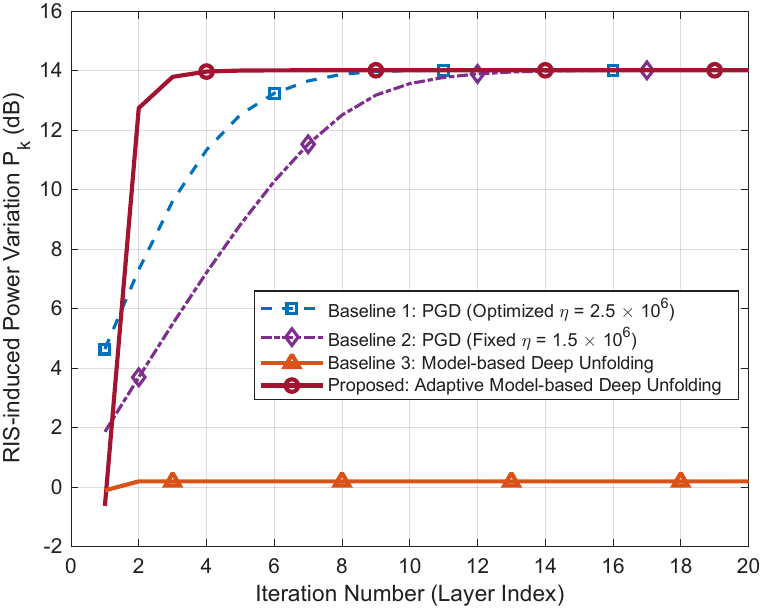}%
    }%
    \hfill
    \subfigure[User located at (7.5, 5.5), Max.\label{fig:iter_7555_max}]{%
        \includegraphics[width=.48\linewidth]{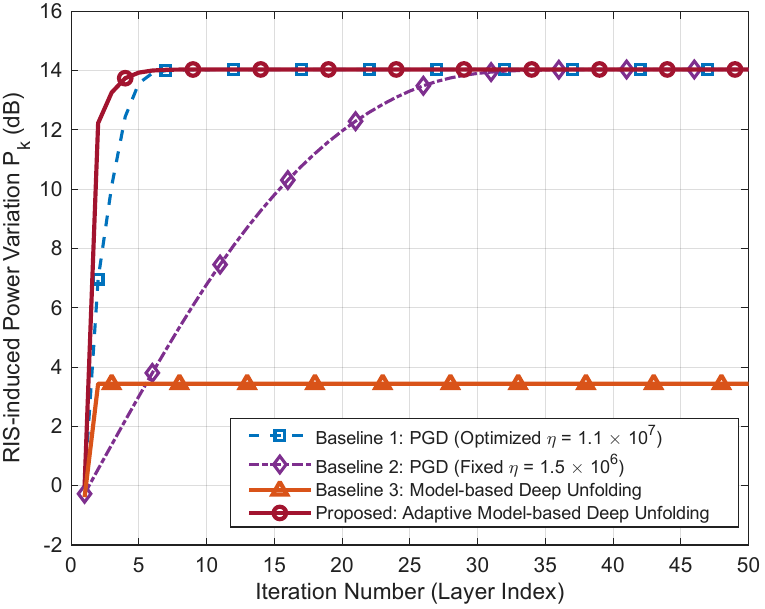}%
    }%

    \caption{Convergence performance comparison of single-user case.}
    \label{fig: single_iter_comparison}
\end{figure}
We next evaluate the proposed AI-based schemes (denoted as \emph{Ours (Single)} and \emph{Ours (Joint)}) in approaching the optimal performance in the single‑UE setting. Specifically, we compare our methods with several benchmark algorithms, including the SDR-based method~\cite{WU19enhance,maliciousRIS}, PGD (with fixed and optimized~$\eta$)~\cite{PGD}, and the baseline model-based DU~\cite{modelAI}. These schemes correspond to the theoretical frameworks and optimization strategies detailed in Section~\ref{sec:baseline}, and Section~\ref{subsec: modelAI}, respectively. 

As shown in Table~\ref{tab:result comparison all locations}, for single‑UE minimization tasks, the PGD method with an optimized~$\eta$ consistently achieves the best performance across all evaluated scenarios. In contrast, the standard SDR approach fails to provide reasonable minimization results (see the discussion in Section~\ref{sec:single scheme}). The baseline model-based DU also exhibits limited effectiveness, as the absence of an appropriate scaling factor~$\lambda$ prevents it from achieving meaningful optimization in minimization tasks. By incorporating the proposed scaling mechanism, our newly enhanced AI‑assisted DU attains a competitive performance that is comparable to PGD approaches. 
For maximization tasks, all methods (except the baseline model-based DU) yield comparable results. 
Meanwhile, the proposed AI‑assisted method is highly efficient computationally, requiring very low execution time. Our current reporting time is mainly limited by the testing platform. Because it is implemented using a neural‑network architecture, its efficiency can be further enhanced through standard parallel computing techniques when hardware permits, making it suitable for large‑scale or real‑time deployment.

As illustrated by the convergence curves in Fig.~\ref{fig: single_iter_comparison}, the proposed method consistently converges faster than the baseline PGD with a fixed step size in both maximization and minimization tasks. Although PGD with an optimized step size can achieve faster convergence and somewhat deeper suppression, the optimal step sizes of the two PGD variants differ substantially, as indicated by the $\eta$ values in the legend. This suggests that PGD is highly sensitive to step-size selection and requires careful manual tuning. In contrast, the proposed adaptive deep-unfolding method automatically learns suitable update parameters during training and therefore avoids manual hyperparameter tuning.


\subsection{Evaluations on Multi-User Optimization Limits}

We then evaluate the achievable performance in multi-user scenarios. To validate the optimization boundaries, a combination of the classic scalarization-based approach (defined in Section~\ref{sec:Boundary determination}) and a Monte Carlo experiment (based on $10^7$ random RIS configurations) are employed. For the two-user case, each random trial produces a unit-norm reflection vector and the corresponding power pair $(P_1, P_2)$. The resulting samples form a point cloud, from which a convex lower envelope can be extracted as a reference boundary.

\begin{figure}[ht]
    \centering
    \subfigure[Monte Carlo cloud and Scalarized Optimization boundaries.\label{fig:suba}]{%
        \includegraphics[width=0.48\linewidth]{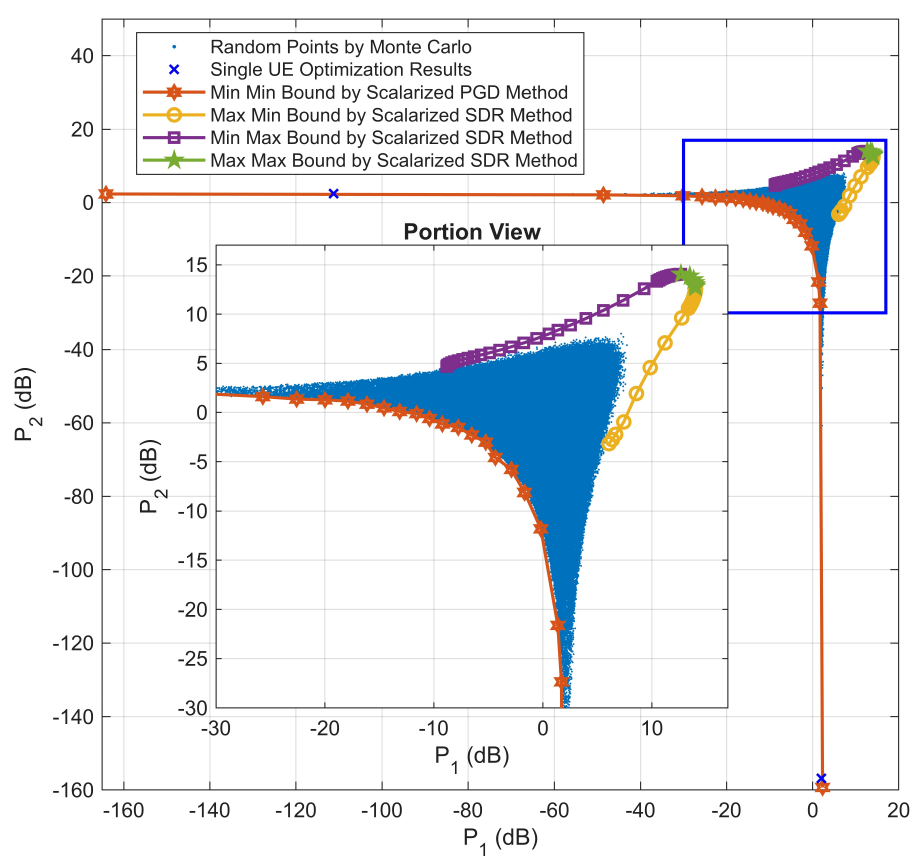}%
    }%
    \hfill
    \subfigure[Zoomed-in optimization boundaries.\label{fig:subb}]{%
        \includegraphics[width=0.48\linewidth]{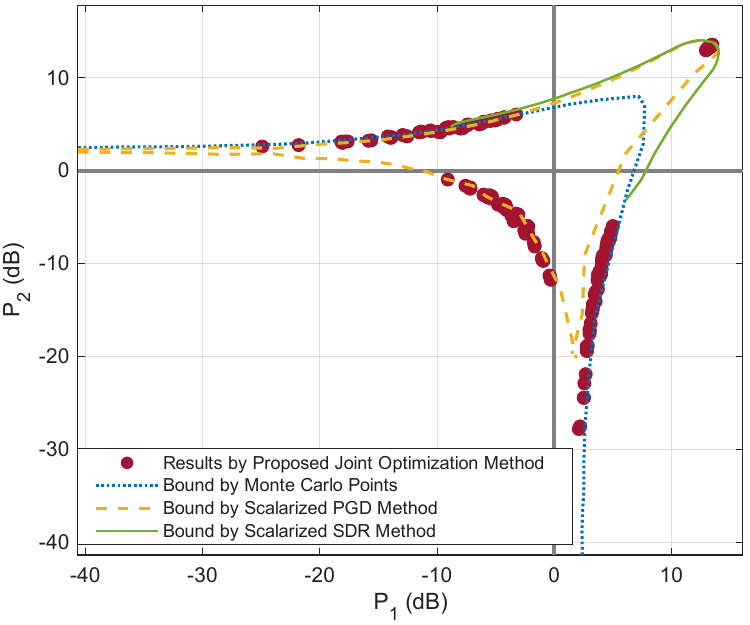}%
    }
    \caption{Monte Carlo cloud and optimization boundaries for two users located at (4.8, 1.3) and (7.5, 5.5).}
    \label{fig:MC_BD_2_1}
\end{figure}

Fig.~\ref{fig:MC_BD_2_1} compares the boundaries obtained from scalarization with those derived from Monte Carlo sampling. The scalarization-based results closely follow the concave hull of the sampled points, especially around the joint minimization region (e.g., $P_1 = -4.34$ dB and $P_2 = -4.58$ dB). This agreement indicates that scalarization provides an effective estimator of the achievable performance frontier. However, pure Monte Carlo exploration becomes unreliable in extreme regions of the boundary due to the rarity of such configurations. This limitation is observed not only in the extreme max–max region but also in single-user extreme points (e.g., minimizing one user's power), where random configurations tend to cluster around the central region of the achievable set.

We also examine two commonly used scalarization-based baselines. First, the scalarized SDR solutions exhibit visible gaps compared with the Monte Carlo envelope, especially in regions where power minimization dominates. This behavior is mainly caused by the relaxation in the SDR formulation, which tends to bias the solution toward amplification-oriented configurations. This behavior is consistent with the discussions in Section~\ref{subsec: all jamming}. Second, the scalarized PGD approach generally provides tighter approximations to the boundary. However, its performance becomes scenario-dependent, and noticeable deviations can still be observed in certain objectives, particularly in the max–min case.

These observations highlight the difficulty of accurately exploring the multi-user RIS power region using individual traditional strategies alone. In contrast, the proposed joint optimization framework consistently approaches the outer boundary of the achievable multi-user power region across all considered scenarios. Compared with the baseline strategies, the proposed method provides stable performance and remains close to the outer boundary under different objective configurations. Moreover, the scalarized trade-off coefficient offers a quantitative mechanism to control the balance between different users, which will be further analyzed in the next subsection.


\begin{figure*}[ht]
    \centering
    \subfigure[$(P_1, P_2)$ 2D projection onto XY-plane.\label{fig: 3UE 12}]{\includegraphics[width=.33\linewidth]{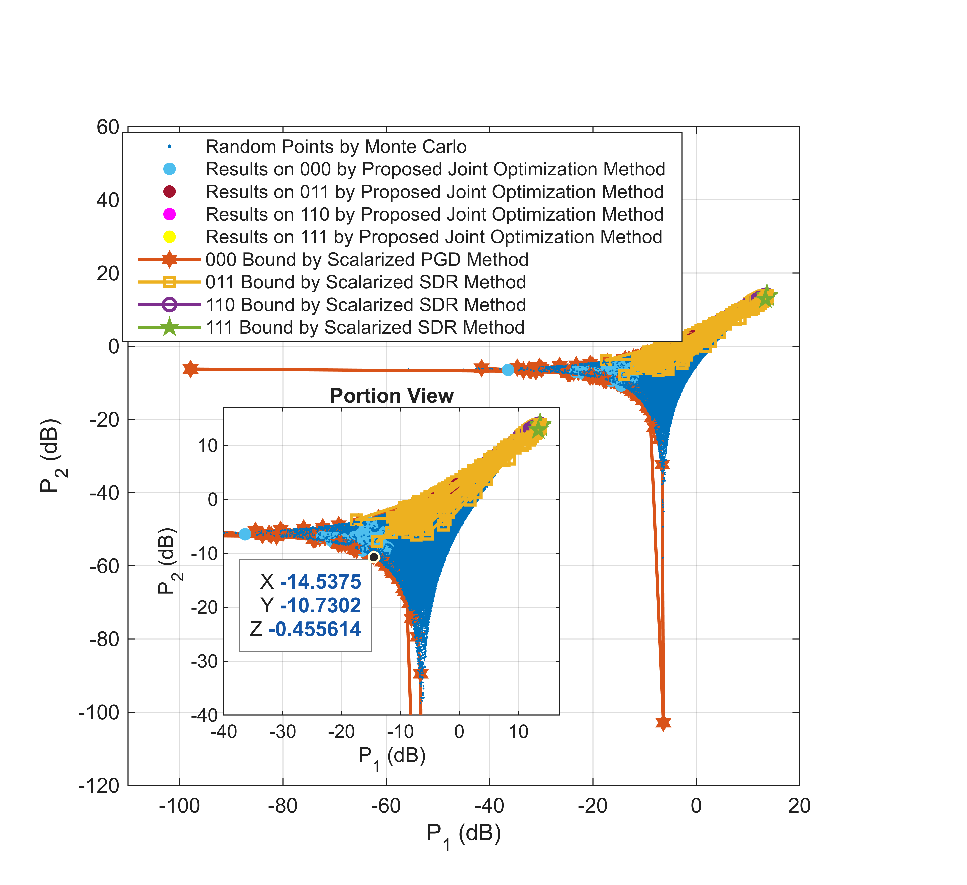}}%
    \subfigure[$(P_1, P_3)$ 2D Projection onto XZ-plane.\label{fig: 3UE 13}]{\includegraphics[width=.33\linewidth]{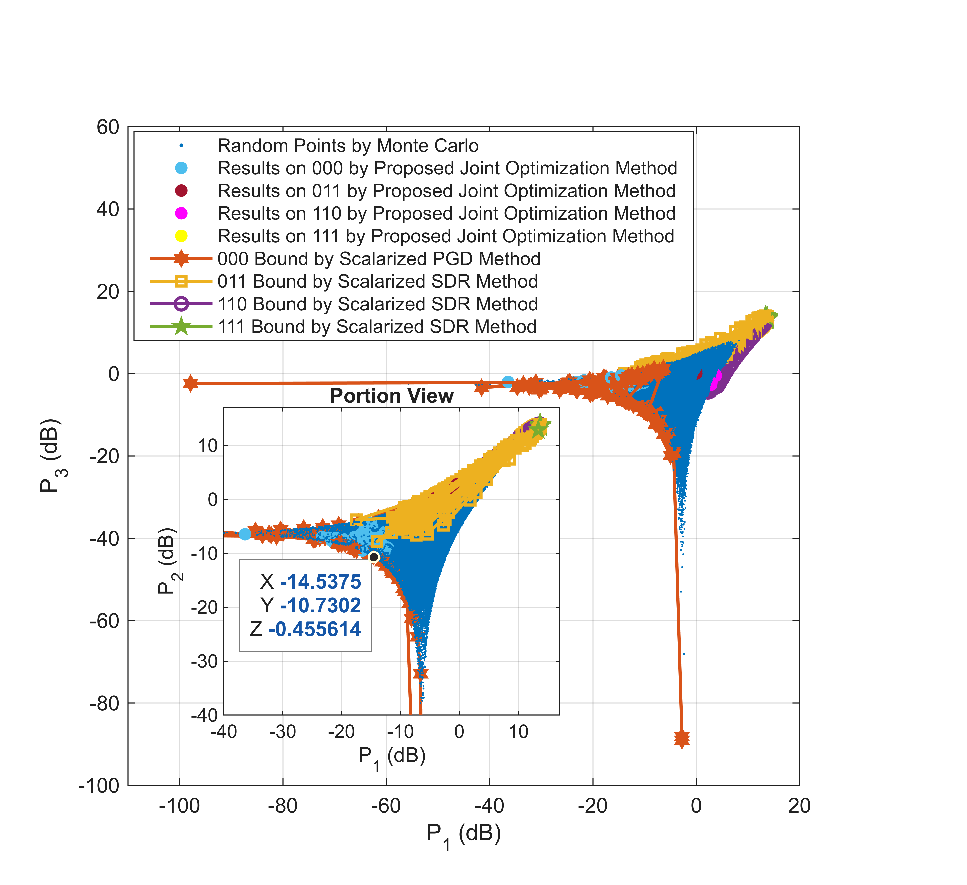}}%
    \subfigure[$(P_2, P_3)$ 2D Projection onto YZ-plane.\label{fig: 3UE 23}]{\includegraphics[width=.33\linewidth]{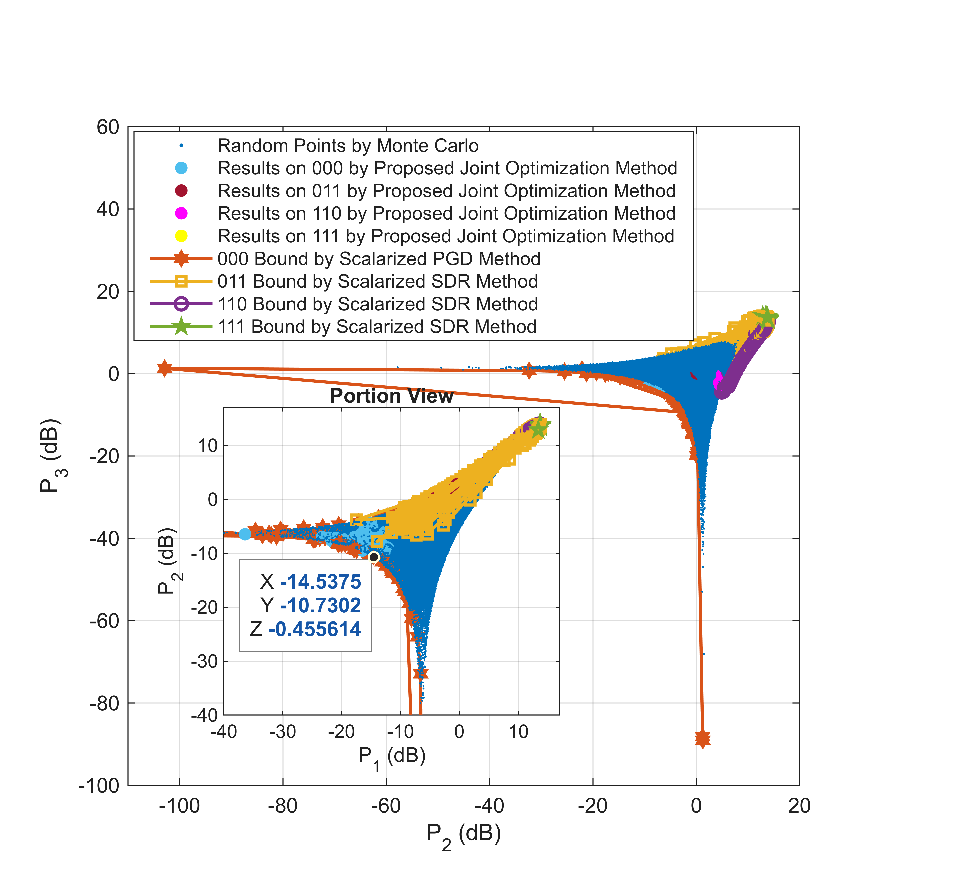}}
\caption{Evaluation results and performance boundaries at $(3, 0.5)$, $(4, 1)$, and $(7, 5.5)$. 
\label{fig: 3UE} }
\end{figure*}

Based on these boundaries, we first examine the two-user scenario. When the RIS attempts to suppress both users simultaneously, the most balanced operating point is approximately $(P_1, P_2) = (-4.34~\text{dB}, -4.58~\text{dB})$, which is significantly weaker than the suppression achievable in single-user settings. This result highlights the inherent trade-off introduced by multi-user interference control. In the signal‑amplification regime (i.e., the max–max scenario, labeled as~11), the maximum power occurs when all reflective paths are perfectly aligned in a common direction. In this case, the received powers converge to nearly identical values $(P_1, P_2) = (13.5 \pm 0.75~\text{dB}, 13.5 \mp 0.75~\text{dB})$.
In the hybrid amplification–suppression scenarios (i.e., the max–min and min–max cases, labeled as~01 and~10, respectively), the linear optimization method described in Section~\ref{subsec: joint jam and amp} is used to determine the boundary. While this method performs poorly for power minimization, it demonstrates strong effectiveness for power maximization. Because maximization dominates the hybrid cases, all RIS elements tend to align in the same direction to enhance the reflected path and surpass the static‑path power. This alignment boosts $P_1$ but also unintentionally increases $P_2$, explaining why the max–min and min–max cases ultimately converge to the same point as the max–max scenario.

We then extend the analysis to the three‑UE case, as illustrated in Fig.~\ref{fig: 3UE}. Since the achievable region becomes three-dimensional, we present two-dimensional projections onto the $(P_1,P_2)$, $(P_1,P_3)$, and $(P_2,P_3)$ planes. The numerical labels in the legends indicate the target configuration for the three‑user setting (e.g., 000 for full suppression, and 011 for suppressing UE~1 while amplifying UE~2 and UE~3). Four representative configurations (000, 110, 011, and 111) are highlighted to demonstrate the RIS’s flexibility in handling asymmetric objectives across multiple users. Although some projected points exhibit $(P_1, P_2)$ values similar to those observed in the two‑user case (e.g., around $-12~\text{dB}$ for both users), it does not imply equally effective suppression for all three users. In most instances, the third user's power is only marginally reduced, revealing the inherent difficulty of extending RIS‑based suppression to multi-user systems. For instance, in Fig.~\ref{fig: 3UE 12}, one such point corresponds to
$(P_1, P_2, P_3) = (-14.54,\ -10.73,\ -0.46)$, indicating minimal suppression for $P_3$.  Similarly, Fig.~\ref{fig: 3UE 13} shows $(P_1, P_2, P_3) = (-9.43,\ -2.04,\ -9.43)$, and Fig.~\ref{fig: 3UE 23} shows $(P_1, P_2, P_3) = (-13.62,\ -7.44,\ -3.27)$. These examples consistently demonstrate that, while balanced suppression may be achieved for two users in a projection plane, the third user generally experiences only moderate suppression. This reflects the fundamental trade-offs inherent in multi‑user RIS optimization. Furthermore, when the RIS attempts to suppress two UEs simultaneously, its ability to provide power amplification deteriorates significantly, underscoring the complexity of handling conflicting objectives in multi‑user environments.

Nevertheless, the results demonstrate that the proposed joint-optimization can approach to the Pareto boundary in all considered scenarios.

\subsection{Joint-Optimization with Varying Element Allocations} \label{subsec: comp on vary element allocation}




\begin{table}[b]
\centering
\caption{Impact of element allocation on joint optimization results for configuration 00 at (4.8, 1.3) and
(7.5, 5.5).}
\label{tab:gt_element_comparison 00}
\begin{tabular}{l|c|c|c|c}
\toprule
 & $P_1$ (dB) & $P_2$ (dB) & UE1  & UE2 \\
\midrule
\multirow{4}{*}{\makecell[l]{$\min\min~(00)$}}
 & -0.6306 & -6.8368 & 46 & 54 \\
 & -2.3521 & -5.0997 & 48 & 52 \\
 & -4.5034 & -2.9117 & 50 & 50 \\
 & -9.8549 & -0.1883 & 52 & 48 \\
\midrule
\multirow{5}{*}{\makecell[l]{$\max\min~(10)$}} 
 & 2.1094 & -21.3769 & 50 & 50 \\
 & 3.0857 & -18.3308 & 54 & 46 \\
 & 4.1538 & -8.9917 & 58 & 42 \\
 & 5.9195 & -2.3967 & 62 & 38 \\
 & 6.6081 & -0.6076 & 66 & 34 \\
\midrule
\multirow{4}{*}{\makecell[l]{$\max\max~(11)$}} 
 & 13.4643 & 13.4584 & 44 & 56 \\
 & 13.4269 & 13.3697 & 46 & 54 \\
 & 13.4198 & 13.3615 & 48 & 52 \\
 & 13.2939 & 13.3020 & 50 & 50 \\
\bottomrule
\end{tabular}
\end{table}

\begin{figure}[ht!]
    \centering
    \includegraphics[width=.75\linewidth]{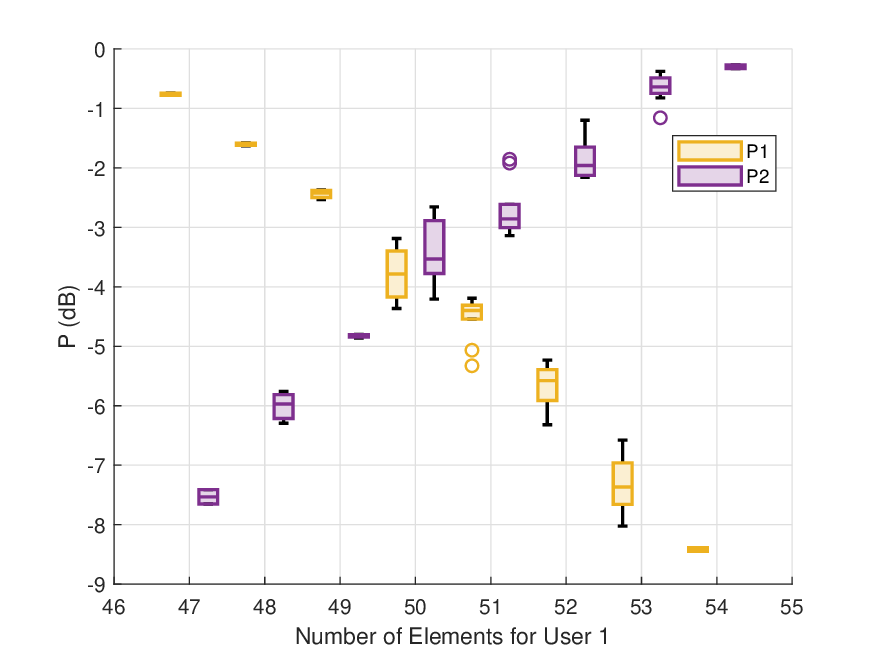}
    \caption{Impact of element allocation on joint optimization results for configuration 00 at  (4.8, 1.3) and (7.5, 5.5).}
    \label{fig: gt_element_comparison 00}
\end{figure}

A key feature of the proposed joint‑optimization method lies in its ability to generate multiple feasible solutions that satisfy all imposed constraints, thereby enabling flexible trade-offs among the UEs. Table~\ref{tab:gt_element_comparison 00} presents several randomly selected allocations for a representative scenario in which UE~1 and UE~2 are located at $(4.8,,1.3)$ and $(7.5,,5.5)$, respectively. Each row in the table corresponds to a feasible solution characterized by a particular allocation of RIS elements between the two users, thus reflecting different user-weighting preferences. The results reveal a narrow operating region in which both users can experience simultaneous power reduction. In the min–min (00) configuration, for example, an allocation of $(46,54)$ (meaning 46 elements for UE~1 and 54 elements for UE~2) strongly prioritizes the suppression of $P_2$, yielding a substantial loss of $-6.83$~dB for UE~2, while UE~1 incurs only a minor loss of $-0.63$~dB. As the allocation shifts slightly to $(52,48)$, the solution transitions toward prioritizing suppression for UE~1, pushing UE~2 close to the suppression boundary with nearly negligible reduction. This sensitivity highlights how fine‑grained control over element distribution can significantly affect the achieved performance and helps pinpoint the precise allocation ratio at which both users begin to benefit concurrently. Fig.~\ref{fig: gt_element_comparison 00} shows detailed outcomes for the 00 configuration. As illustrated, even when the number of RIS elements assigned to each UE is equal, the resulting power losses can differ markedly. This discrepancy arises from unequal path loss due to different user distances from the RIS. Although the proposed joint‑optimization method provides practical flexibility for multi‑user power control and interference management, achieving consistent suppression across multiple UEs with a single RIS panel remains a fundamentally challenging task.

In comparison, the results for the max–min configuration (10) reveal a considerably broader feasible region. A balanced allocation of $(50,50)$ leads UE~1 to reach its amplification boundary, whereas an allocation of $(66,34)$ drives UE~2 toward its suppression boundary. This asymmetry illustrates that achieving mixed objectives, such as maximizing the performance of one user while suppressing the other, often requires a disproportionately uneven distribution of RIS elements.
For the max–max configuration (11), the trade-off between the two users becomes negligible. Nearly all allocations yield favorable outcomes for both users. This behavior is attributed to the dominance of the reflected channel over the static channel in this regime. When more RIS elements are directed toward a particular user, the resulting enhancement of the reflected path increases the total received power, thereby benefiting both users simultaneously.

\subsection{More Discussions on Multi-User Cases}

\begin{table}[ht]
\centering
\caption{Performance (in dB) comparison for different numbers of UEs at arbitrarily selected locations.}
\label{tab:numUE_positions}
\begin{tabular}{l|c|c|c|c|c}
\toprule
 & \#UE & \textbf{(3, 0.5)} & \textbf{(4, 1)} & \textbf{(4.8, 1.3)} & \textbf{(7.5, 5.5)} \\ 
\midrule
\multirow{3}{*}{\makecell[l]{Min all} }
& 1 & -42.4770 & -73.6943 & -91.4964 & -101.4131 \\
& 2 & -12.558  & -22.212  & -21.7359 & -10.2614  \\
& 4 & -9.9589  & -9.6389  & -7.3205  & -1.2062   \\
\midrule
\multirow{3}{*}{\makecell[l]{Max all}} 
& 1 & 13.7845  & 13.9397  & 14.0156  & 14.0377   \\
& 2 & 13.6175  & 13.9397  & 14.0098  & 13.6356   \\
& 4 & 13.6856  & 13.7600  & 13.7407  & 13.4057   \\
\bottomrule
\end{tabular}
\end{table}

Building on the performance guarantees of the joint optimization method, additional multi-user scenarios are explored to assess the effectiveness of a single RIS implementation. Table~\ref{tab:numUE_positions} summarizes the observed power values at different UE locations under varying numbers of simultaneously served UEs, while maintaining a fixed total RIS panel size. The rows represent the number of users concurrently served by the RIS, and the columns correspond to specific user positions.
A clear degradation in signal suppression performance is observed as the number of users increases, indicating that a single RIS panel struggles to maintain effective signal suppression in multi-user settings. In contrast, amplification performance remains nearly unchanged, suggesting that the RIS retains its ability to enhance signals even when serving multiple users.
To ensure a fair comparison across configurations, each setup maintains the same number of RIS elements per user. Specifically, the single-user case assigns 25 elements to one user; the two-UE case activates 50 elements in total (25 per user); and the four-UE case uses all 100 elements, allocating 25 to each user. This design isolates the impact of joint optimization under a fixed per-user element budget.
For each location, the reported power value reflects the best achievable performance across all permutations of the given number of users. For the single-user and four-UE cases, the selection of the best permutation is straightforward. In contrast, the two-UE case involves multiple possible pairings, and the table entries represent the optimal outcome among them. Among all possible pairs, the one achieving the best amplification or suppression performance is selected. This selection strategy ensures that the most representative and competitive results are reported for each configuration. Meanwhile, a more balanced performance can be expected when a fixed UE pair is considered. For instance, the value -12.558 at location (3,0.5) is obtained when this user is paired with (4,1), while the values at (4,1) and (4.8,1.3) are achieved when these two locations are paired with each other. Similarly, the value 
-10.2614 at (7.5,5.5) arises from pairing with (3,0.5). These examples illustrate that each reported value in the 2-UE configuration is not tied to a single fixed pair, but instead reflects the best pairing achievable for that specific location.

To contextualize the suppression performance against a CSI-free baseline,
Table~\ref{tab:disco_comparison} compares the proposed framework with a
DISCO-inspired random passive RIS baseline~\cite{DISCO_TWC25, DISCO_MWC},
which applies random time-varying phase configurations to degrade UE performance
without requiring any target CSI. Two variants are evaluated using the same
Sionna RT channels and $P_k$ metric: \textit{DISCO Mean} (average over $10^5$
random 1-bit configurations) and \textit{DISCO Best} (oracle best-of-$10^5$
trials). As shown, DISCO Mean produces near-zero or positive $P_k$ values
($+0.013$ to $+0.143$~dB), confirming that CSI-free random jamming is entirely
ineffective for instantaneous power suppression, while even the oracle DISCO Best
achieves only $-1.3$ to $-3.1$~dB --- the fundamental ceiling of any CSI-free
random approach. In contrast, the proposed method achieves $-42.5$ to
$-101.4$~dB for $K=1$ and $-1.2$ to $-10.0$~dB for $K=4$, representing a
gain of \textbf{more than 40~dB} over DISCO Best, quantitatively confirming
that the CSI-aware optimization provides a suppression capability fundamentally
unattainable by CSI-free random methods.

\begin{table}[t]
\centering
\caption{Comparison with DISCO-style CSI-free random baseline~\cite{DISCO_TWC25,
DISCO_MWC} for Min~all objective ($P_k$ in dB, $N=100$, $10^5$ random trials).}
\label{tab:disco_comparison}
\resizebox{\columnwidth}{!}{%
\begin{tabular}{clcccc}
\toprule
\textbf{\#UE} & \textbf{Method} &
\textbf{(3,\,0.5)} & \textbf{(4,\,1)} &
\textbf{(4.8,\,1.3)} & \textbf{(7.5,\,5.5)} \\
\midrule
\multirow{3}{*}{$K=1$}
 & DISCO Mean & $+0.013$ & $+0.046$ & $+0.048$ & $+0.017$ \\
 & DISCO Best & $-1.332$  & $-1.940$  & $-2.861$  & $-1.837$  \\
 & \textbf{Ours} & $\mathbf{-42.477}$ & $\mathbf{-73.694}$
                 & $\mathbf{-91.496}$ & $\mathbf{-101.413}$ \\
\midrule
\multirow{3}{*}{$K=2$}
 & DISCO Mean & $+0.063$ & $+0.064$ & $+0.067$ & $+0.061$ \\
 & DISCO Best & $-2.221$  & $-2.827$  & $-3.082$  & $-2.835$  \\
 & \textbf{Ours} & $\mathbf{-12.558}$ & $\mathbf{-22.212}$
                 & $\mathbf{-21.736}$ & $\mathbf{-10.261}$ \\
\midrule
\multirow{3}{*}{$K=4$}
 & DISCO Mean & $+0.143$ & $+0.139$ & $+0.141$ & $+0.140$ \\
 & DISCO Best & $-1.762$  & $-3.107$  & $-2.337$  & $-1.497$  \\
 & \textbf{Ours} & $\mathbf{-9.959}$  & $\mathbf{-9.639}$
                 & $\mathbf{-7.321}$  & $\mathbf{-1.206}$  \\
\bottomrule
\end{tabular}%
}
\end{table}

In summary, our extensive numerical evaluations across various configurations yield the following key observations:
\begin{itemize}
    \item Efficiency in Signal Enhancement: In single-UE maximization tasks, while all compared methods achieve comparable power gains, our proposed AI-assisted framework delivers these results with significantly lower computational latency, making it more suitable for real-time deployment.
    \item Effectiveness in Signal Suppression: Traditional SOTA methods, particularly SDR and baseline model-based DU without scaling, consistently perform poorly in signal suppression due to numerical instability or lack of step-size adaptation. Our method effectively addresses this bottleneck, achieving suppression depths comparable to finely-tuned iterative benchmarks without requiring manual parameter intervention.
    \item Stability and Transparency in Multi-User Scenarios: While scalarized optimization can be sensitive to weight selection, our joint-optimization framework maintains robust performance near the Pareto boundary. Furthermore, the introduced element allocation strategy renders the competitive trade-offs between users transparent, providing a quantifiable measure of priority-driven resource contention.
    \item Scalability Limits of Single RIS: As the number of simultaneously served users increases, a noticeable degradation in signal suppression is observed. This underscores the inherent physical limitations of using a single RIS panel for multi-user coordination, where enhancement objectives remain robust while suppression capability scales poorly with user density.
\end{itemize}

\section{Conclusion}\label{sec:conclusion}

This work exploited the achievable performance limit of optimizing a single RIS panel in multi-user networks with potentially conflicting objectives.To efficiently explore this design space, we first introduce an adaptive gradient-scaling mechanism that stabilizes and accelerates the convergence of the underlying optimization process across varying channel and system conditions, reducing the need for manual hyperparameter tuning. Building on this foundation, the proposed framework partitions the RIS panel based on dynamic user priorities and employs modular element adjustment to enable fine-grained control over signal enhancement and suppression while maintaining high flexibility. A low-complexity beamformer recovery method further reduced computational overhead in single-user cases. The approach was evaluated across three scenarios-full amplification, full suppression, and hybrid configurations, using boundary characterization and large-scale Monte Carlo validation, which confirmed the accuracy and near-optimality of the proposed method. Extensive simulations in realistic channel conditions demonstrated that the framework consistently outperformed SDR-based baselines, highlighting its potential as a practical solution for next-generation wireless systems requiring real-time multi-user trade-off management. Future work includes extending the framework to imperfect CSI settings, where the modular 
per-user optimization structure naturally supports robust optimization extensions without 
redesigning the overall architecture.

\bibliographystyle{IEEEtran} 
\bibliography{reference}

\end{document}